\documentclass[prb]{revtex4}
\usepackage{epsf}
\begin{document}
\title{Collapsing supra-massive magnetars: FRBs, \\ the repeating FRB121102 and GRBs}

\author{Patrick Das Gupta and Nidhi Saini}
\affiliation{Department of Physics and Astrophysics, University of Delhi, Delhi - 110 007 (India)}
\email{pdasgupta@physics.du.ac.in}





\begin{abstract}
Fast radio bursts (FRBs) last for $\sim $ few milli-seconds and, hence, are likely to arise from the  gravitational collapse of supra-massive, spinning neutron stars after  they lose the centrifugal support (Falcke \& Rezzolla 2014). In this paper, we provide arguments to show that the repeating burst, FRB 121102,  can also be modeled in the collapse framework provided the supra-massive object implodes either into a Kerr black hole surrounded by highly magnetized plasma or into a strange quark star. Since the estimated rates of FRBs and SN Ib/c are comparable, we put forward a common progenitor scenario for FRBs and long GRBs in which only those compact remnants entail prompt $\gamma $-emission whose kick velocities are almost  aligned or anti-aligned with the stellar spin axes. In such a scenario, emission of detectable   gravitational radiation and, possibly, of neutrinos are expected to occur during the SN Ib/c explosion as well as, later, at the time of magnetar implosion.
\end{abstract}


\maketitle


\section{Introduction}
Ever since the serendipitous discovery of FRB 010724, the very first fast radio burst (FRB) gleaned from archival pulsar survey  data by 
Lorimer and his team members (Lorimer {\it et al.} 2007), about 24  FRBs    have been  detected so far whose physical nature still continue to confound astrophysicists (Katz 2016; Zhang 2017). FRBs are bright radio transients with very high brightness temperatures, lasting for $\sim $ few milliseconds with peak flux densities ranging from $\sim 0.1 $ Jy to $ \sim 10$ Jy at frequencies of about $\sim $ 700 MHz - 2 GHz.

Associated large dispersion measure (DM) $\sim 500 \ - \ 1200 \ \mbox{pc}\ \mbox{cm}^{-3}$ and high Galactic latitudes strongly suggest that  FRBs are extragalactic events with redshifts $z$ in the range $ \sim 0.3 $ to $ \sim 1$, implying that they lie at distances $\gtrsim $ 1 Gpc, if DM is largely due to electrons in the IGM. As of now, polarization data exist  only for few FRBs like FRB 110523 ($z < 0.5)$, FRB 140514 ($z \lesssim 0.5$) and FRB 150215 (luminosity distance $< 3.3$ Gpc), with the first showing 44 \% linear polarization, second less than 10 \% linear polarization but $\sim $ 20 \% circular polarization and the third,  about 43 \% linear polarization and 3 \% circular polarization (Petroff {\it et al.} 2015; Ravi \& Lasky 2016; Petroff {\it et al.} 2017).

Have FRB counterparts been seen in other wavebands? FRB 131104, the first one to be detected in a targeted search using Parkes radio telescope, lies in the direction of Carina dwarf spheroidal galaxy (Ravi {\it et al.} 2015). According to DeLaunay {\it et al.} (2016), the gamma ray transient Swift J0644.5−5111 is a counterpart of FRB 131104 at 3.2 $\sigma $ confidence level. The estimated DM for this  FRB is about $779\  \mbox{pc \ cm}^{-3} $ implying a redshift  $z \cong 0.55$ while  Swift J0644.5−5111 has a fluence and duration of about $4 \times 10^{-6} \ \mbox{erg \ cm}^{-2}$ and 377 s, respectively  (Ravi {\it et al.} 2015; DeLauney {\it et al.} 2016;  Murase  {\it et al.} 2017). However, Shannon \& Ravi (2017) report that absence of radio afterglow in the direction of Swift J0644.5−5111 strongly constrains the energetics indicating that this gamma ray transient is unlikely to be a standard long duration gamma ray burst (GRB).

 If FRBs result from such catastrophic events then a natural model that can account for their millisecond duration  radio emission at $\sim $ 1 GHz  is  that of  initially rapidly rotating supra-massive neutron stars collapsing into black holes (BHs) due to loss of centrifugal support  as they spin down due to magnetic braking (Falcke \& Rezzolla 2014; Zhang 2013). 
 
  However, a  catastrophic event like a collapsing neutron star is unlikely to produce a FRB candidate that exhibits  recurring outbursts. There certainly is one such source from which  sporadic  radio transients are observed - FRB 121102 (Spitler {\it et al.} 2016; Scholz {\it et al.} 2016).  About 200 such  intermittent outbursts  in radio   have already been detected from it so far.  This repeater is unlikely to be an active magnetar since recent simultaneous observations  of FRB 121102 with the help of {\it Chandra Observatory} and {\it XMM-Newton} place stringent upper limits,  $3 \times 10^{-11} \ \mbox{erg} \  \mbox{cm}^{-2}$ on X-ray fluence during the episodic radio-bursts and  $3 \times 10^{41} \ \mbox{erg} \  \mbox{s}^{-1}$  on a possible persistent X-ray luminosity (Scholz {\it et al.} 2017). 
  
Several models for such repeating FRBs have been   posited ranging from  intense plasma wind sweeping across the magnetosphere of an extragalactic pulsar (Zhang 2017),  episodic relativistic $\mbox{e}^\pm$  wind from an active galactic nucleus (AGN) impinging on a plasma cloud nearby (Vieyro {\it et al.} 2017),  active young remnant of a neutron star or a magnetar (Kashiyama \& Murase 2017; Metzger   {\it et al.} 2017), intense flares from young magnetars leading to shock induced maser emission (Kulkarni {\it et al.} 2014; Lyubarsky 2014; Beloborodov 2017) to asteroids falling randomly on a neutron star (Dai {\it et al.} 2016; Bagchi 2017).

In the present study, we explore two distinct scenarios to show that a repeating FRB can also be brought within the ambit of  a collapsing supra-massive magnetar framework. In the penultimate  section, we also delve in  the possibility of linking FRBs with long GRBs.

\section{Magnetar Collapse and FRB 121102}

Magnetars are highly magnetized neutron stars (NSs) with surface $|\vec {B} | \gtrsim  10^{14}$ Gauss. However,   they spin relatively at slower rates (period $P \approx $ 2 - 12 s) compared to pulsars. They  display  episodic  intense X-ray outbursts  that are powered by  internal  magnetic field instead of  rotational kinetic energy (Thompson \& Duncan 1993; Usov 1993). Magnetars could have originated initially with high spin rates ($P \approx $ few milliseconds) so that   small seed magnetic fields could have been amplified to  very high $B$ by $\alpha - \Omega $ dynamo action,  and then, because of magnetic dipole radiation depleting the rotational kinetic energy, their period  increased steadily (Thompson \& Duncan 1993). 

It is plausible that if the core of a rotating, massive ($M_* \gtrsim 35\ M_\odot$)  Wolf-Rayet like star   collapses eventually to form a rapidly spinning  ($P \sim 1 $ millisecond), supra-massive magnetar ($M_0 \gtrsim 3\ M_\odot$), then as this remnant spins  down due to magnetic braking, it can  further collapse   after it loses the centrifugal support (CS). In particular,
since the observed radio transients from FRBs have millisecond duration, the model posited by Falcke \& Rezzolla (2014) attains a special status,  in which FRBs result from  gravitational collapse of spun down supra-massive NSs  on  dynamical time scales,
\begin{equation}
\tau_{coll} \sim \sqrt{\frac{R^3} {G M}}\sim 1/\sqrt{G \rho_{nuc}}\ \sim 10^{-3} \ \mbox{s ,}
\end{equation}
where $M$ and $R$ are the  mass and radius, respectively, of the NS at the onset of the collapse, while $\rho_{nuc}\cong 10^{12} - 10^{14} \ \mbox{gm\ cm}^{-3}$ is the nuclear density inside the NS.

If the NS collapses to a final radius $R_f$,  energy that can be released is given by,
\begin{equation}
\Delta E \cong G M^2 \ \bigg (\frac {1} {R_f} - \frac {1} {R} \bigg )
\end{equation}
According to the above expression, in case the NS collapses to form a black hole (BH),   energy generated  and shared among  high energy particles like photons and neutrinos created during the implosion  is of supernova explosion  energy scale,
\begin{equation}
\Delta E \approx 8 \times 10^{53}\ \mbox{erg\ ,}
\end{equation} 
(One has substituted  $M=1.4 \ M_\odot$,   $R=12$ km  and $R_f=2 G M/c^2$ = 4.2 km in eq.(2), to obtain eq.(3). It will be an order of magnitude higher if $M \sim 2.5 \ M_\odot$.)

However, several factors like short time scale of collapse, small mean free path  due to density being $>\rho_{nuc}$ and ever increasing space-time curvature prohibit  the hot, relativistic  particles to escape  out of the collapsing object. Under such conditions,  only  the radiation  caused by physical processes that take place in the ephemeral magnetosphere  left behind outside   is likely to be  detected,  entailing an event akin to a FRB (Falcke \& Rezzolla 2014). Since the deduced FRB brightness temperatures are $\gtrsim 10^{30}\ {}^\circ $K, one needs to invoke coherent emission mechanisms to explain these  observed bright, short duration radio transients. 

For many FRBs, flux density $S_\nu$ shows power law behavior, i.e.  $S_\nu \propto \nu^{\ \alpha} $. The isotropic luminosity $L$ for FRBs then can readily be obtained from the expression,
\begin{equation}
L=6.76\times 10^{43} \ \bigg (\frac {\nu} {1.4\ \mbox{GHz}}\bigg )^{1+\alpha} \bigg (\frac {S_\nu} {2\ \mbox{Jy}}\bigg )\bigg (\frac {D_L} {10\ \mbox{Gpc}}\bigg )^2\ \ \mbox{erg} \  \mbox{s}^{-1}
\end{equation}
where $D_L$ is the luminosity distance corresponding to the cosmological parameters $\Omega_m=0.27$, $\Omega_\Lambda=0.73$ and $H_0=68$ km/s/Mpc. Given millisecond duration, eq.(4) implies that FRBs radiate energy $\sim 10^{38}\ - \ 10^{41}$ erg in radio, about 12 orders of magnitude less compared to the total energy released in most stellar core collapse events. 

Falcke and Rezzolla (2014) have shown that it is possible to obtain the required radio luminosity (eq.(4)) from  curvature radiation  in the magnetosphere wherein  bunched charge particles, $\sim N_{bunch} $ in number,   flow out coherently with a net velocity    along  smoothly bent (radius of curvature $R_B$) magnetic field lines  emitting energy at a rate given by,
\begin{equation}
L_{curv} = \frac {2 \gamma^4 N^2_{bunch} e^2  c} {3 R^2_B}
\end{equation}
\begin{equation}
\ \ \ \ \ \approx 2 \times  10^{43}\ \bigg (\frac {N_{bunch}} {2\times 10^{28}} \bigg )^2 \bigg (\frac {\gamma} {10^2} \bigg )^4 \bigg (\frac {R_B} {30\ \mbox{km}} \bigg )^{-2} \ \mbox{erg} \ \mbox{s}^{-1}
\end{equation}
 at a  characteristic radio frequency,
\begin{equation}
\nu_{curv}= \frac {3 \gamma^3 c} {4 \pi R_B} \approx 2.4 \ \bigg (\frac {\gamma} {10^2} \bigg )^3 \bigg (\frac {R_B} {30\ \mbox{km}} \bigg )^{-1} \ \mbox{GHz} 
\end{equation}
where $\gamma $ is the Lorentz factor of the charge particles. 

The magnetic field lines that were originally anchored to the NS matter are expected to last only for $\sim $ few milliseconds, the time scale over which the magnetar collapses to acquire a newly formed event horizon (EH).
But this scenario, as it stands, obviously cannot explain recurring radio transients detected from FRB 121102.

It has been discovered recently that FRB 121102 is located in a dwarf galaxy  at $z\cong 0.193$ that  also contains a  compact radio-source (Chatterjee {\it et al.} 2017; Tendulkar {\it et al.} 2017). This  persistent radio-source is relatively weaker with $S_\nu \sim 200\ \mu\mbox{Jy}$ while the flux densities at 1.7 GHz   from FRB 121102, on different occasions, vary from $\sim 0.1\ \mbox{Jy}$ to $\sim 4\ \mbox{Jy}$. The compact radio-source and the FRB  are separated by a projected distance less than $\sim $ 40 pc (Marcote {\it et al.} 2017).

The question whether the only known repeating radio burst belongs to an altogether different class of FRBs has also been addressed in the literature. Some advocate that all FRBs are repeaters and it is just the higher sensitivity of Arecibo radio telescope that could pick up the intermittent radio transients from FRB 121102 (Lu \& Kumar 2016; Spitler {\it et al.} 2016).

On the other hand, for the currently available FRB sample, Palaniswamy \& Zhang (2017) studied the distribution of ratios of adjacent peak flux densities  and the time intervals between corresponding successive bursts and  concluded that either the repeater belongs to a distinct class or that all FRBs  physically originate from a common progenitor system with FRB 121102, however,  being extra-active.  
Espousing their latter conclusion, we attempt here to reconcile the catastrophic model involving magnetar collapse with the  observed repeater. For this purpose, we explore two distinct scenarios  - one that involves Blandford-Zjanek process and the other, related to  compact stars made of strange quark matter.
\subsection{Collapse to a Kerr BH and Blandford-Zjanek process}
\vskip 1 em
 Suppose, apart from a co-rotating magnetosphere within the light-cylinder, the magnetar that imploded  to give rise to the repeating FRB is surrounded by magnetized plasma. After losing the CS, this spun down magnetar  collapsed into a Kerr BH. Since FRB 121102 lies close to a weak AGN in a dwarf galaxy (Chatterjee {\it et al.} 2017; Marcote {\it et al.} 2017; Tendulkar {\it et al.} 2017), one may argue  that the episodic AGN wind triggers sporadic accretion of the magnetized matter present outside its EH. Because of the   Blandford-Znajek mechanism (BZM) this leads to an unsteady bipolar jet emanating from  FRB 121102 (Blandford \& Znajek 1977).

As the jet drills through the ambient plasma, it can cause  shocked shells to develop and, as discussed by Waxman (2017) recently (in the context of  AGN itself driving the shocked shells  in the dwarf galaxy), can entail emission of coherent radiation at GHz frequency by virtue  of a synchrotron maser action, provided  that the number density of electrons in the shocked plasma, as a function of the Lorentz factor $\gamma_e$, increases faster than $\gamma^3_e$. 

In what follows, we make quick estimates of jet energy resulting from the BZM. The rotating EH of a Kerr BH embedded in a plasma  threaded with $\vec B$ is, in many ways, analogous to a rotating conducting shell in an external magnetic field (Hartle 2003). Hence, the potential difference  generated due to changing magnetic flux is given by,
\begin{equation}
V_{el} \sim \frac {1}{c} (\Omega_H/2 \pi)(B. \pi R^2_{BH})
\end{equation}
where $\Omega_H=\frac {J} {M R_s R_{BH}} $ is the angular speed of the EH. 
The EH radius of the Kerr BH is,
\begin{equation}
R_{BH}=\frac {R_s} {2} + \sqrt{\bigg (\frac {R_s} {2} \bigg )^2 - \bigg (\frac 
{J}{Mc} \bigg )^2}
\end{equation}
where $R_s\equiv\frac {2 G M} {c^2} $, and $M$ is mass of the BH remnant.

The ensuing jet power due to  mining of BH rotational energy via BZM is given by (Hartle 2003), 
\begin{equation}
L_{B-Z} \approx \mbox{current} \times V_{el} \sim   \frac {V^2_{el}}{\mathcal{R}} \sim \frac {c\ V^2_{el}}{4}
\end{equation}
For extremal BHs, $J=cMR_s/2 \ \Rightarrow \ R_{BH}=R_s/2 $ and $\Omega_H=\frac {c}{R_s} \ \Rightarrow \ V_{el} \sim R_s B/8$, so that,
\begin{equation}
L_{B-Z}\approx \frac {c B^2 R^2_s}{256}
\end{equation}
\begin{equation}
\ \ \ \ =5.45 \times 10^{43} \bigg ( \frac {B} {10^{12}\ \mbox{G}} \bigg )^2\ \bigg (\frac {M} {2.3\ M_\odot} \bigg )^2 \ \mbox{erg} \ \mbox{s}^{-1}
\end{equation}
From the above equation it is evident that if the synchrotron maser, driven by the jet-induced shocked shells (Waxman 2017), radiates at a rate that is $\sim 0.1 $ fraction of the jet luminosity, one can explain the recurring radio-transients. 

However, few caveats are in order:
\vskip 1 em

(a) Eq.(12) requires B $\sim 10^{12}$ G to be  present near the EH of the  BH remnant. When a magnetar with polar B field $\sim  10^{14}\ - \ 10^{15}$ G collapses to form a BH, can an ambient plasma threaded with magnetic field $\sim 10^{12}$ G be maintained near the  EH? (In the context of GRB modeling, Contopoulos {\it et al.} (2017) find that $B \sim 10^4 $ G can hover near the horizon.)

(b) Accretion of plasma by itself cannot generate such high $\vec {B}$  since the Eddington limit for the magnetic field is,
\begin{equation}
B_{Edd}=\sqrt{\frac {8 \pi c^4 m_p}{\sigma_T G M}}\cong 2 \times 10^8 \ (M/2.3\ M_\odot)^{-1/2}\ G \ .
\end{equation}

The required high $\vec {B}$ could also arise from a magnetized  nebula around the magnetar that collapsed to form the Kerr BH. For instance, a  wind nebula has been discovered near the magnetar Swift J 1834.9-0846 (Younes et al. 2016; Granot {\it et al.} 2016).

We now move on to another scenario that is relevant when the magnetar is much more massive and, therefore, has to rotate faster in order to garner the required CS.

\subsection{Formation of a Strange Star from the Magnetar Collapse} 
\vskip 1 em
It is evident from eq.(9) that the angular momentum of the BH cannot exceed $cMR_s/2=GM^2/c$. In that case, what happens if the core of the magnetar which collapses has angular momentum too large to form a BH? 

Suppose, 
 the initial angular momentum of the supra-massive magnetar is $J_0 \sim M_0 R^2_0 \Omega_0$, with  initial angular speed $\Omega_0 < \Omega_{Break-up} \equiv \sqrt{\frac{G M_0}{R^3_0}}$ so that the magnetar does not get torn apart due to centrifugal force. 
As the magnetar spins down due to  magnetic braking, it starts  contracting so that at any instant of time $t$, $ M_0 R^2 (t) \Omega(t) \sim \beta J_0$, assuming that there is no mass loss during  the process. (The factor $\beta \lesssim 1$ has been included because a small fraction of the initial angular momentum  gets carried away by the magnetic dipole radiation.)

As the magnetic braking proceeds, the equatorial region separates from the rest of the contracting magnetar, forming a disc-like structure because of  centrifugal force being larger there, while the central core of mass $M_c=M_0 - M_{disc}$ and angular momentum $J_c=\beta J_0 - J_{disc}$ collapses gravitationally.
Taking the disc to be sufficiently thin so that angular momentum per unit mass is $\sqrt{GM_c r}$ (assuming local orbital speed to be Keplerian) one obtains,
\begin{equation}
J_{disc}=\int^{r_D} _{R_c} {\sqrt{GM_c r}\  \Sigma (r)\ 2\pi r dr}
\end{equation}
\begin{equation}
\ \ \ \ \approx 4 \pi \rho_{disc} h \int^{r_D}_{R_c}{\sqrt{GM_c r} \  r dr}
\end{equation}
where we have assumed that the disc density $\rho_{disc} (r) $ and the disc thickness $h (r) $ do not vary  appreciably between the size $R_c$ of the core and the outer disc radius $r_D$. 

One can estimate the outer radius $r_D$ of the disc by assuming that it is the equatorial radius of the contracting magnetar when the equatorial centrifugal force  just crossed the gravitational force so that $\Omega=\sqrt{\frac{G M_0}{r^3_D}}$. Therefore,
\begin{equation}
J=\beta J_0=M_0 r^2_D \Omega=M_0\sqrt{G M_0 r_D}\Rightarrow r_D=\frac {\beta^2 J^2_0}{G M^3_0}
\end{equation}
Since $M_{disc} \approx \pi h \rho (r^2_D - R^2_c)$, it can be easily shown from eq.(15) that,
\begin{equation}
J_{disc}=\frac {8}{5} M_0 (1 - M_c/M_0) \sqrt{GM_c r_D} \  \bigg ( \frac {1- (R_c/r_D)^{5/2}}{1-(R_c/r_D)^2} \bigg )
\end{equation}


From the extremal BH criteria that follows from eq.(9), no BH is  formed if $J_c = \beta J_0 - J_{disc} > \frac {GM^2_c} {c}$. Then, making use of eqs.(16) and (17), the condition for no BH formation can be expressed as,  
\begin{equation}
\beta J_0 > \frac {GM^2_c}{c} +  \frac {8}{5} M_0 (1 - M_c/M_0) \sqrt{GM_c r_D} \  \bigg ( \frac {1- (R_c/r_D)^{5/2}}{1-(R_c/r_D)^2} \bigg )
\end{equation}

\begin{equation}
\ \ \ = \frac {GM^2_c}{c} +  \frac {8}{5} \beta J_0 (1 - M_c/M_0) \sqrt{\frac{M_c} 
{M_0}}\  \bigg ( \frac {1- (R_c/r_D)^{5/2}}{1-(R_c/r_D)^2} \bigg )
\end{equation}
so that one arrives at,
\begin{equation}
J_0 > \frac {GM^2_c}{c \beta [1 - \frac {8} {5} (1-M_c/M_0)(M_c/M_0)^{1/2}  \frac {1- (R_c/r_D)^{5/2}}{1-(R_c/r_D)^2} ]}
\end{equation}
as a condition for no BH formation.

 What is the fate of such a spinning and collapsing core? As it contracts, its density rises and when the density overshoots $\sim (5\ - \ 7) \times 10^{14} \mbox{gm\ cm}^{-3}$, it turns into a strange star (SS), composed  roughly of equal number of u, d,  s quarks and a small fraction of electrons, to maintain charge neutrality  (Itoh 1970; Witten 1984; Alcock {\it et al.} 1986). During the collapse, a burst of thermal photons and neutrinos with average energy $\sim $ few MeV is  emitted from the SS. As long as the total baryon number is $\gtrsim 100$, this strange matter is more stable compared to  the standard hadronic matter. 
 
Mass-radius relation for SSs goes as $M\propto R^3$ so that (Xu 2005),
\begin{equation}
R\cong 10\ \bigg (\frac {M} {M_\odot} \bigg )^{1/3}\ \mbox{km}
\end{equation}
This may be contrasted with that of the NS case (Lattimer \& Schutz 2005), 
\begin{equation}
R \cong 15 \ \bigg ( \frac {M} {2.3 \ M_\odot} \bigg )^{1/4}  \ \mbox{km}
\end{equation}


As an illustration, if one substitutes  in eq.(20),  $\beta = 0.8$, $M_0=2.3  \ M_\odot $ and $M_c = 1.4 \ M_\odot $, one finds  $\Omega_0 > 5 \times 10^3 \ \mbox{rad\  s}^{-1}$ in order that no BH is formed. Eq.(21) has been used to obtain the value for $R_c$, and $r_D=15$ km has been assumed (this is the radius of a NS of mass $2.3 \ M_\odot$).
 One may note that this lower limit on the initial angular speed does not violate the initial stability criteria since for $M_0=2.3  \ M_\odot $, one can still have $5 \times 10^3 \ \mbox{rad\  s}^{-1} < \Omega_0 < \Omega_{Break-up}\sim \sqrt{\frac{G M_0}{r^3_D}} = 9.5 \times 10^3 \ \mbox{rad\  s}^{-1}$. So, in this alternate scenario for FRB 121102, a supra-massive magnetar, spinning very rapidly could have collapsed into a rotating SS.
 

Since the quarks inside the SS are held together by strong forces, the electrons due to their mutual electrostatic repulsion (and no strong interactions) migrate radially outwards,  distributing themselves above the quark matter   surface within a height of $\sim $ few 100 fm .  This results in a strong  radially outwards electric field $|\vec {E}| \approx 5\times 10^{17}\ - \ 10^{18}$ V/cm that keeps  this  thin layer of electrons bound to the SS (Alcock {\it et al.} 1986).

One can readily estimate the number of electrons $N_e$ in the layer above the SS using Gauss law,
\begin{equation}
N_e=\frac{4 \pi}{e} |\vec {E}| R^2 \cong 7 \times 10^{36}\ \bigg (\frac {|\vec {E}|} {10^{18}\ \mbox{V/cm}} \bigg ) \bigg (\frac {M} {M_\odot} \bigg )^{2/3}
\end{equation}
Eq.(21) has been made use of in  the above equation.

As FRB 121102 lies close to a persistent radio-source, sporadic flaring up of the AGN can drive a $\mbox{e}^\pm$  wind that triggers various oscillatory modes of the electron layer  of the SS. We  make simple estimates to arrive at the power radiated due to dipole oscillation. If $N_{dp}$ denotes the effective number of electrons that participate in dipole oscillation stimulated by the AGN wind, the  induced electric dipole  moment is $d_e (t) \sim e N_{dp} (R+\delta Z (t))$ (for simplicity one has assumed a z-direction deformation of the electron layer). Then, 
\begin{equation}
\dot {d_e} \sim e N_{dp} \dot{\delta Z} \Rightarrow \ddot {d_e} \sim e N_{dp} \ddot{\delta Z}= e N_{dp} \bigg (\frac {e |\vec E |} {m_e} \bigg )= \frac{e^2 N_{dp}|\vec E |}{m_e}
\end{equation}
so that the luminosity $\frac {2 |\ddot{\vec {d_e}}|^2} {3  c^3}$ due to dipole radiation is given by,
\begin{equation}
L_{dip}   \approx \frac {2 e^4 N^2_{dp}|\vec E |^2} {3 m^2_e c^3}=4\times 10^{43}\ \bigg ( \frac{N_{dp}}{10^{14}}\bigg )^2 \bigg ( \frac {|\vec E |} {10^{18}\ \mbox{V/cm}} \bigg )^2 \ \mbox{erg} \ \mbox{s}^{-1}
\end{equation}
Hence, only a minuscule fraction of the total number of electrons $N_e$ (eq.(23)) needs to participate in  the dipole oscillation to generate the required power.
Using detailed calculations,  Mannarelli {\it et al.} (2014) have shown that torsional oscillation of the thin electron layer relative to the positively charged SS can cause emission of GHz radiowaves with luminosity as high as $\sim 10^{45}$ erg/s.
If FRB 121102  indeed is associated with a rapidly spinning SS then due to r-mode instability  one would  expect the repeater  to be also a  persistent source of gravitational waves (Andersson, Jones \& Kokkotas 2002).

\section{GRBs and FRBs: Common Progenitor System?}

It is well established that long duration GRBs   ($T_{90} >$ 2 s) are associated with type Ib/c supernovae and thus, are likely to be   the byproduct of collapsing  Wolf-Rayet  stars (for recent reviews, see Granot {\it et al.} 2015; Kumar \& Zhang 2015). Prompt $\gamma $-emission ensues from internal shocks in colliding shells ejected from a central engine  along narrow jets with opening solid angle $\Delta \Omega \sim $ 0.01 str (i.e. $\theta_{jet} \sim 0.1 $ radian). Observed afterglows reveal interesting interplay of relativistic shocks and radiative processes (Reshmi \& Bhattacharya 2008). Hard X-ray polarimetry of AstroSat has succeeded recently in detecting prompt emission polarization in the case of  GRB 151006A, even though it is a faint burst lasting for $\sim $ 20 s (Rao {\it et al.} 2016). 

Short GRBs ($T_{90} <$ 2 s), on the other hand, constitute  only $\sim $ 30\% of the total GRB population and are plausibly  the aftermath of final merger of binary NSs. However, a common progenitor model for  short and long GRBs based on jet opening solid angles in the ratio $\sim $ 3:7 was  speculated upon (Das Gupta 2004). The sub-pulses of short GRBs   show distinct statistical correlations with other observed parameters (Bhat, Gupta \& Das Gupta 2000; Das Gupta 2000; Gupta,  Das Gupta \& Bhat a, b 2000).

It is interesting to note that the rate of  SN Ib/c  is $\sim 6 \times 10^4 \ \mbox{Gpc}^{-3} \mbox{yr}^{-1}$ (Frail  {\it et al.} 2001) and  from all sky rate $\sim 2100 \ \mbox{day}^{-1}$  for FRBs brighter than $\sim 2$ Jy the ensuing FRB rate is 
 $\sim 10^4 - 10^5 \ \mbox{Gpc}^{-3} \mbox{yr}^{-1}$ (Champion {\it et al.} 2016). While SN Ib/c rate and FRB rate are of comparable magnitude,  long GRB rate has been estimated to be  $\sim 10^2 - 10^3 \ \mbox{Gpc}^{-3} \mbox{yr}^{-1}$ (i.e. $\sim  10^{-3} - 10^{-2}$ times the FRB rate). 

Motivated by the remarkable similarity in the FRB and SN Ib/c rates, we briefly explore a scenario in which all FRBs and majority of long GRBs are associated with the collapse of the central iron core of WR stars. FRB 121102 too lies inside a low-metallicity dwarf galaxy and such galaxies tend to host hydrogen-deficient, high luminosity supernovae and long GRBs (Lunnan {\it et al.} 2014; Scholz {\it et al.} 2017).  On the other hand, in the context of FRB-GRB connection, Zhang (2013) had proposed that a large fraction of GRBs would lead to FRBs after $10^2 \ - \ 10^4$ s, and as a consequence, some FRBs would be  physically associated with GRBs.

{\it Swift}/XRT observations have revealed  X-Ray plateau followed by steep decay in  many long GRBs that could naturally be explained by magnetic braking of a rapidly spinning magnetar (Metzger {\it et al.}, 2011; Granot {\it et al.} 2015).
Adopting this basic picture for long GRBs, we posit that   
FRBs and long GRBs  result from the collapse of massive ($\gtrsim 35 \ M_\odot$), rotating WR stars. 
The unstable Fe core, having mass $\sim 2-3\ M_\odot $ and initial size $\sim 10^8 $ cm, of such a  rotating He star would undergo free fall on a time scale of $\sim 1 $ s, giving rise to a rapidly  spinning supra-massive proto-neutron star (SPNS)  that moves with a  velocity $\vec v_{kick} $ ranging from 0 to $\sim $ 1000 km/s with respect to the progenitor star because of a pre-natal kick. The SPNS would then cool and contract to form a spinning supra-massive NS/magnetar  in $\sim 5-10 $ s.

Due to the rapid rotation and the sudden  collapse of the core, two narrow funnel shaped regions oriented  along the stellar spin axis, pointing away from each other and straddling across a near spherical cocoon of size $\sim 10^8$ cm surrounding the SPNS,  are created (Meszaros \& Rees 2001). These regions are  essentially devoid of  stellar  (baryonic) matter.
If the direction of $\vec v_{kick} $ lies within the funnel shaped regions, the impellent gamma ray jet ensuing  later from the magnetar would not face any baryon loading problem  and thereby could break out of the He envelope, giving rise to a long GRB. It is to be noted that WR stars  are relatively small helium stars with size $\sim \mbox{few}  \times 10^{10} $ cm, as they have already blown their upper H layer  away.
 
If the opening angle $\theta_F$ of the funnel shaped regions of the WR star is also  $\sim $ 0.1 rad, corresponding to a solid angle $\Delta \Omega_F \sim 0.01 $ str, then the probability that the prompt gamma rays are not quenched due to baryon loading is,
\begin{equation}
\approx 2 \frac {\Delta \Omega_F} {4 \pi} \approx 5 \times 10^{-3} \ \ ,
\end{equation}
which is consistent with long GRB rate being $\sim  10^{-3} - 10^{-2}$ times the FRB rate.

If the SPNS moves in any other direction it would encounter  stellar material after traversing the cocoon, and due to baryon loading the $\gamma$-rays resulting from electron-positron pairs generated by the neutrino wind would get degraded to softer photons and so, there would be no prompt $\gamma$-emission.  Even with a low kick velocity,  $v_{kick}\sim 100 $ km/s,  the SPNS would cross the baryon evacuated cocoon of size $\sim $ 1000 km in $\sim $ 10 s, well  before the compact remnant becomes  transparent to the neutrinos. In fact, magnetars with high kick velocities  have been observed. For instance, the kick velocity component perpendicular to the line of sight of magnetar Swift J 1834.9-0846 is $\sim 580 $  km/s provided it is at a  distance  $\sim $ 15 kpc (Younes et al. 2016; Granot et al. 2016).

In other words, the SPNS could lead to a GRB  with a probability $\sim 10^{-3}$, assuming that  $\theta_F \sim $  0.1 rad (eq.(26)), if its kick velocity is more or less aligned or anti-aligned with the rotation axis of the progenitor star. In all other situations, the compact remnant will be hindered by  the stellar matter before the emergence of the $\nu-\bar{\nu}$ wind. The spinning magnetar would continue to accrete matter as it crosses the He envelope until the supernova explosion occurs. Eventually, the supra-massive magnetar would collapse after it loses the CS, giving rise to  a burst of radio emission due to the physical processes taking place in the temporary magnetosphere outside the EH. In this scenario, one would expect emission of gravitational waves with frequency in the $\sim $ kHz band and, possibly of  neutrinos with energy $\sim $ few MeV, twice - first,  during the SN Ib/c event and  second time, when the magnetar thereafter collapses to form a BH. A detailed study on this subject is under progress.

\section{Conclusion}
Implosion of spun down supra-massive NSs can naturally lead to FRBs. 
In this paper, we have shown that it is possible to explain recurring  bursts from FRB 121102 by  invoking either  sporadic  Blandford-Znajek process occurring near a Kerr black hole  or  torsional oscillations of electrons close to the surface of a strange star, both triggered by AGN wind from the weak radio-source  co-located in the dwarf galaxy. If the observed radio transients are attributed to electronic oscillations  of the rotating strange star, one would expect to detect persistent  gravitational waves from the compact remnant due to its r-mode instability.

Since  the FRB and the SN Ib/c rates  are almost equal, it is plausible that the collapsing iron core of a WR star gives rise to a moving supra-massive magnetar (due to an imparted pre-natal kick) whose implosion shows up as a FRB in most cases and, in those rare situations, when its kick velocity is nearly either aligned or anti-aligned with the rotation axis of the star, the spinning down magnetar launches prompt emission of $\gamma$-ray bursts before it implodes.  If this picture is right, one would expect to see emission of observable gravitational waves and of neutrinos, first  during the SN Ib/c event and also later,  during the radio burst.



\section*{Acknowledgement}
PDG thanks Dr. Samir Mandal, Dr. Santabrata Das and other co-organizers of RETCO-III for an  enriching academic experience  at IIST, Thiruvananthapuram.
PDG also takes pleasure in thanking Professors J. Granot, Pawan Kumar, A. R. Rao, Alak Ray, Dipankar Bhattacharya and Poonam Chandra for fruitful discussions and a stimulating atmosphere during the workshop on Gamma Ray Bursts held at NCRA, Pune. 



\end{document}